\documentclass[%
 reprint, 
 amsmath,amssymb,
 aps,
floatfix,
longbibliography
]{revtex4-1}

\usepackage{enumitem}
\usepackage{graphicx}
\usepackage{qcircuit}
\usepackage{tikz}

\usepackage{bm}
\usepackage{xcolor}
\usepackage[colorlinks=true,citecolor=blue,linkcolor=magenta]{hyperref}

\DeclareMathOperator{\tr}{tr}

\begin{document}

\title{Recovery of Damaged Information and the Out-of-Time-Ordered Correlators}

\author{Bin Yan}
\affiliation{Center for Nonlinear Studies, Los Alamos National Laboratory, Los Alamos, New Mexico 87545}
\affiliation{Theoretical Division, Los Alamos National Laboratory, Los Alamos, New Mexico 87545, USA}
\author{Nikolai A. Sinitsyn}
\affiliation{Theoretical Division, Los Alamos National Laboratory, Los Alamos, New Mexico 87545, USA}

\date{\today}

\begin{abstract}
  The evolution with a complex Hamiltonian generally leads to information scrambling.
 A time-reversed dynamics unwinds this scrambling and thus leads to the original information recovery.
 We show that if the scrambled information is, in addition, partially damaged by a local measurement, then such a damage can still be treated by application of the time-reversed protocol. This information recovery  is described by the long-time saturation value of a certain out-of-time-ordered correlator  of local variables.  
We also propose a simple test that distinguishes between  quantum and reversible classical chaotic information scrambling.
\end{abstract}

\maketitle

In complex strongly correlated systems, local information  spreads quickly over the whole system.  This process is characterized by exponentially fast changes of the out-of-time ordered correlators (OTOCs) \cite{Larkin1969,Kitaev2015,Maldacena2016-mb,Swingle2018-xe}, such as the following  correlator of local operators $W$ and $V$ with specific time ordering:
\begin{equation}
    F(t) = \langle W^\dag(t)V^\dag(0)W(t)V(0)\rangle.
\end{equation}
After OTOCs saturate, initially local information becomes encoded into global entanglement, hindering the data from local measurements.

 It can be  hard to recover this information if the scrambling path is not completely known or  if the final state is partly damaged.
 For instance, a single qubit  thrown into a black hole is quickly dispersed and lost behind the horizon. With the resource of early Hawking radiation, only a few qubits of information emitted from the black hole are needed to reconstruct the lost qubit \cite{Hayden2007-cc} but there is no simple recipe for how to do this  without considerable knowledge about the system \cite{Landsman2019-oi,Vermersch2019-fy,Yoshida2019-uq}.


In order to suggest a solution to similar problems,  we consider a practically accessible scenario for information scrambling and unscrambling.
Let us describe this scenario as a hypothetical application of a quantum processor, such as the one in the quantum supremacy test \cite{Qsupremacy}, for hiding quantum information. Our processor can be simpler than that in Ref.~\cite{Qsupremacy} because we require that only one of the qubits can be prepared and measured, which is suitable for experiments with  liquid-NMR quantum computers  \cite{Li2017-il,Samal2011-sm}. 

Let Alice have such a processor that implements fast information scrambling during a reversible unitary evolution of many interacting qubits. 
She applies this evolution to hide an original state of one of her qubits, which we call the central qubit. 
The other qubits are called the bath.
To recover the initial central qubit state, Alice can apply a time-reversed protocol. 

\begin{figure}[t!]
    \centering
    \includegraphics[width=0.45\textwidth]{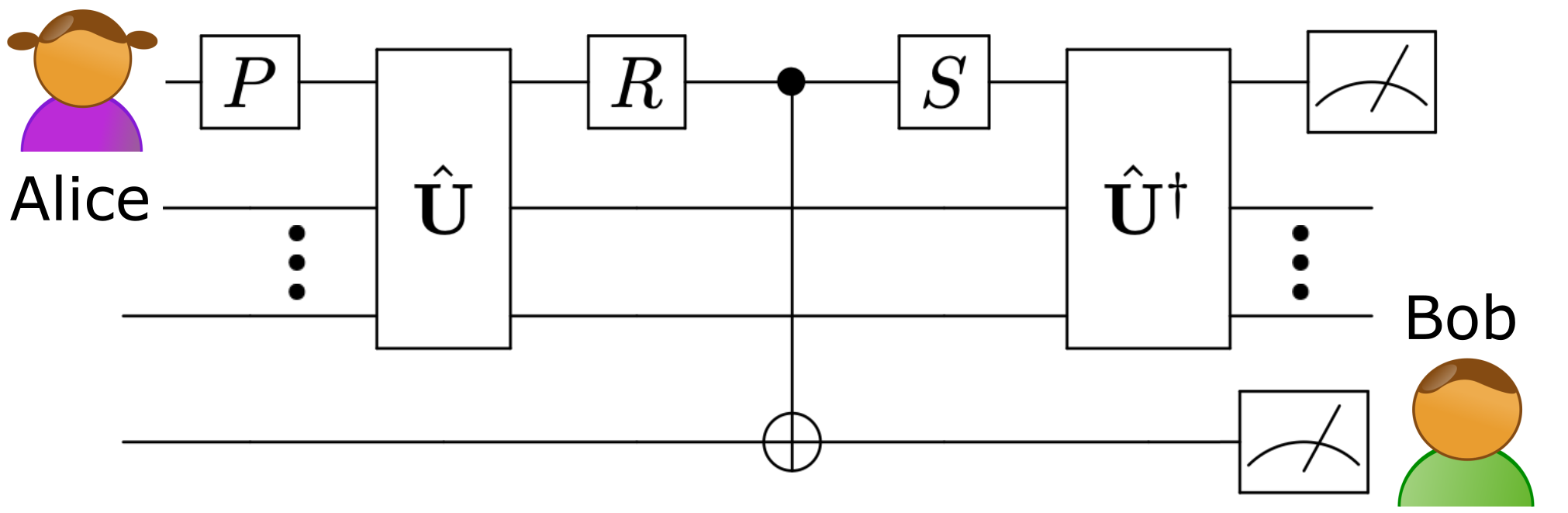}
    \caption{The protocol: Alice prepares the central qubit with the gate $P$, and applies the scrambling unitary $\hat{U}$. Bob measures the central qubit in any basis defined by the gate $R$, with $S$ flipping the qubit to the corresponding post measurement state. Alice is still able to reconstruct the encoded information via a single decoding unitary $\hat{U}^\dag$.}
    \label{fig:device}
\end{figure}

Let Bob be an intruder who can measure the state of the
central qubit in any basis unknown to Alice, as shown in Fig.~\ref{fig:device}. If her processor has already scrambled the information, Alice is sure that Bob cannot get anything useful. However, Bob's measurement changes the state of the central qubit and also destroys all quantum correlations between this qubit and the rest of the system. According to the no-hiding theorem \cite{Braunstein2007-kw}, information of the central qubit is completely transferred to the bath during the scrambling process. However,  Alice does not have knowledge of the bath state at any time. How can she recover the useful information in this case?

In this Letter, we show that even after Bob's measurement, Alice can recover her information by applying the time-reversed protocol and performing a quantum state tomography with a limited amount of effort. Moreover, reconstruction of the original qubit will not be influenced by Bob's choice of the measurement axis and the initial state of the bath.

This effect cannot be explained with semiclassical intuition. Indeed, classical chaotic evolution magnifies any state damage exponentially quickly, which is known as the butterfly effect. The quantum evolution, however, is linear. This explains why, in our case, the uncontrolled damage to the state is not magnified  by the subsequent complex evolution. Moreover, the fact that Bob's measurement does not damage the useful information follows from the property of entanglement correlations in the scrambled state \cite{Braunstein2007-kw}.
Hence, the information recovery effect can be used in practice to distinguish quantum scrambling from the scrambling achieved via classical chaotic dynamics. 

Suppose that the process in Fig.~\ref{fig:device} is realized in a system of qubits (spins-$1/2$'s). The  system starts from an initial product state, $\rho_0=|i\rangle\langle i|\otimes\rho_B$. Here, the central qubit state $|i\rangle$ encodes the information to be scrambled and recovered, whereas $\rho_B$ of the bath qubits can be an arbitrary pure or mixed state. 

After a unitary evolution during time $t_1$, a projective measurement along a random axis is applied to the central spin, without collecting any data from the measurement outcome. Then, the system evolves backward in time during $t_2$, followed by a state tomography for the central spin. We claim that when $t_2=t_1$, the final measurements contain information that can be used to fully reconstruct the initial state of the central qubit.

Two remarks are in order. First,  the recovery of damaged information is a generic effect insensitive to the detailed dynamics, as long as the unitary evolution is complex enough to scramble the information. Thus, for any initial state, after the scrambling process the reduced density matrix of the central qubit becomes maximally mixed. For instance, the desired scrambling unitary can be generated by a chaotic many-body Hamiltonian, or be comprised by random two-qubit unitaries on a quantum circuit. 
Second, the initial information is fully recovered at the moment $t_2=t_1$, i.e., right after the backward unitary becomes conjugated to the forward unitary, but it will be useful to explore what happens for $t_2\ne t_1$.

Let $\hat{P}_r$ be the projection operator for Bob's measurement on the central qubit. There are two complementary histories in which, after Bob's measurement, the central spin state is projected to the subspace described by either $\hat{P}_r$ or $\hat{\mathbb{I}}-\hat{P}_r$. The probability for Alice's projective measurement $\hat{P}_f$ at the final time moment is 
\begin{equation}\label{sum-prob}
        \text{Prob}(\hat{P}_f)=\int [dr]\ \text{Prob}(\hat{P}_f,\hat{P}_r)+\text{Prob}(\hat{P}_f,\hat{\mathbb{I}}-\hat{P}_r),
\end{equation}
where $\text{Prob}(\hat{P}_f,\hat{P}_r)$ is the joint probability  that both Alice and Bob find unit measurement outcomes for their measurement operators. Note that, since we do not collect results of the Bob's intermediate measurements, his complementary measurement outcome contributes to $\text{Prob}(\hat{P}_f)$ as well.
The integral over $r$ in Eq.~(\ref{sum-prob}) accounts for averaging over an arbitrary distribution of possible directions for Bob's measurement axes. 

Let $\hat{U}(t_1)$ be the evolution operator for the time-forward protocol during time $t_1$ and $\hat{U}^\dag(t_2)$ be the evolution operator for the time-reversed protocol during time $t_2$. 
The probability of the nonzero outcome for the intermediate measurement $\hat{P}_r$ and the corresponding post measurement state $\rho_r$ are then given by
\begin{equation}
\begin{aligned}
   & \text{Prob}(\hat{P}_r) = \tr \hat{P}_r\rho(t_1)\hat{P}_r,\\
   & \rho_r = \hat{P}_r\rho(t_1)\hat{P}_r/\text{Prob}(\hat{P}_r),
\end{aligned}
\end{equation}
where $\rho(t_1)=\hat{U}(t_1)\rho_0 \hat{U}^\dag(t_1)$ is the system state at time $t_1$. The probability for the final measurement $\hat{P}_f$, conditioned on the system being projected to the post measurement state $\rho_r$, is
\begin{equation}
        \text{Prob}(\hat{P}_f|\hat{P}_r) =  \tr \left(\hat{P}_f \hat{U}(t_2) \rho_r \hat{U}^\dag(t_2) \hat{P}_f \right).
\end{equation}
This gives the desired joint probability $\text{Prob}(\hat{P}_f,\hat{P}_r) = \text{Prob}(\hat{P}_f|\hat{P}_r) \text{Prob}(\hat{P}_r)$. Since the  system is initially in the state $\rho_0=|i\rangle\langle i|\otimes\rho_B$, the joint probability can be expressed in a compact form:
\begin{equation}\label{eq:correlatorP}
    \text{Prob}(\hat{P}_f,\hat{P}_r)=\langle \hat{P}_r(t_1)\hat{P}_f(t_1-t_2)\hat{P}_r(t_1)\hat{P}_i\rangle.
\end{equation}
Here, the ensemble average is defined as $\langle \bullet \rangle \equiv \tr (\bullet\  \hat{\mathbb{I}}\otimes\rho_B)$. Equation~(\ref{eq:correlatorP})  shows that the effect of Bob's interference on the information that Alice obtains after applying the time-reversed protocol is described by a two-time OTOC of projection operators. 

Let us now express this OTOC  (\ref{eq:correlatorP}) in terms of Pauli operators, using the identity $\hat{\sigma}_\phi \equiv 2\hat{P}_\phi-\hat{\mathbb{I}}$. We are interested in long scrambling times and $t_2
\sim t_1$. The second order spin correlators, such as $\langle\hat{\sigma}_f(t_1)\hat{\sigma}_i\rangle$, decay quickly with time.
Hence, we can safely neglect all such correlators except the ones that depend on $t_1-t_2 \ll t_1$. The joint probability is then
\begin{equation}\label{eq:jointP}
\begin{aligned}
        \text{Prob}(\hat{P}_f,\hat{P}_r)=&\frac{1}{4}+\frac{1}{16}\langle \hat{\sigma}_f(t_1-t_2)\hat{\sigma}_i \rangle+\\
        &\ \ + \frac{1}{16}\langle \hat{\sigma}_r(t_1)\hat{\sigma}_f(t_1-t_2)\hat{\sigma}_r(t_1)\hat{\sigma}_i \rangle.
\end{aligned}
\end{equation}
For $t_1=t_2$, the second term on the right-hand side in Eq.~(\ref{eq:jointP}) is independent of the evolution unitary. All such details are hidden in the third term.  At $t_1=t_2$, this four-point correlator becomes a standard spin OTOC, i.e, 
\begin{equation}
F(t)=\langle \hat{\sigma}_r(t) \hat{\sigma}_i \hat{\sigma}_r(t) \hat{\sigma}_f \rangle.
\label{otoc-spin}
\end{equation}
For finite $t$ and for a small bath this correlator has a nontrivial system-specific behavior that obscures the contribution of $\langle\hat{\sigma}_f(t_1-t_2)\hat{\sigma}_i \rangle$ in Eq.~(\ref{eq:jointP}).  However, we are interested in the typical complex unitary evolution that scrambles information. 
Hence, we claim that $F(t)$ saturates to a universal value that is described by its average over an ensemble of random unitaries. This average can be evaluated as an integral over all unitaries with respect to the Haar measure \cite{Cotler2017-oi,Yan2019-ff}, i.e., 
\begin{equation}\label{eq:Haaraverage}
        \bar{F} = \int_{Haar} dU\  \tr \left[  U^\dag\sigma_rU \sigma_i U^\dag\sigma_rU \sigma_f \rho_B \right],
\end{equation}
where $\rho_B$ is the initial state of bath qubits. The integral can be further calculated using the identity for Haar unitaries \cite{Vermersch2019-fy}:
\begin{equation}
\begin{aligned}
    &\overline{U_{m_1n_1}U^*_{m'_1n'_1}U_{m_2n_2}U^*_{m'_2n'_2}}\\
    = & \frac{\delta_{m_1m'_1}\delta_{m_2m'_2}\delta_{n_1n'_1}\delta_{n_2n'_2}+\delta_{m_1m'_2}\delta_{m_2m'_1}\delta_{n_1n'_2}\delta_{n_2n'_1}}{N^2-1}\\
     - & \frac{\delta_{m_1m'_1}\delta_{m_2m'_2}\delta_{n_1n'_2}\delta_{n_2n'_1}+\delta_{m_1m'_2}\delta_{m_2m'_1}\delta_{n_1n'_1}\delta_{n_2n'_2}}{N(N^2-1)},
\end{aligned}
\end{equation}
where $N$ is the dimension of the Hilbert space. 
After summing over all indices and using a trivial identity ${\rm tr}(\hat{\sigma}_{f,i,r})=0$, the average reduces to
\begin{equation}
    \bar{F} = \langle\sigma_i\sigma_f\rangle/(N^2-1) \equiv \tr(\sigma_i\sigma_f\otimes\rho_B)/(N^2-1).
\end{equation}
We need this formula only for $N\gg 1$ because  then a single typical unitary  produces  the effect that coincides with the average of the OTOC.
The large denominator,  for $N\rightarrow \infty$, makes this OTOC decay to zero. Since this happens for random unitary evolution, the same is true for sufficiently long scrambling  times and sufficiently large baths, so the fourth order correlator in Eq.~(\ref{eq:jointP}) also vanishes. 

After averaging over random unitaries, the joint probability $\text{Prob}(\hat{P}_f,\hat{P}_r)$ is the same as $\text{Prob}(\hat{P}_f,\hat{\mathbb{I}}-\hat{P}_r)$, so the final probability is twice that of $\text{Prob}(\hat{P}_f,\hat{P}_r)$ for any distribution of Bob's measurement axes.
Equation~(\ref{eq:jointP}) reduces then to
\begin{equation}\label{eq:finalP}
    \text{Prob}(\hat{P}_f)=\frac{1}{2}+\frac{1}{8}\langle \hat{\sigma}_f(t_1-t_2)\hat{\sigma}_i \rangle.
\end{equation}
This is the main result of our work. It shows that Alice's measurement probability at $t_2=t_1$ can be used to construct the initial state of the central qubit, i.e., $\rho_i = (\hat{\mathbb{I}}+\hat{\sigma}_i)/2$. This information returns to the central qubit during a short central spin lifetime, as it is described by a second order spin correlator. The state of the central qubit at $t_2=t_1$ is
\begin{equation}
\label{rho-fin}
\rho_f(t_2=t_1)=\frac{1}{4}\hat{\mathbb{I}}+\frac{1}{2}\rho_i.
\end{equation}
Note that the central qubit  ends up in a partially mixed state.  However, distinct initial states are transformed to distinct final states, i.e., the map is injective. Hence, the initial state can be extracted from the final state  statistically, using quantum state tomography. Importantly, this can be done with any precision using only a finite ensemble of our system because the coherent contribution in Eq.~(\ref{rho-fin}) is not suppressed by the  complexity of the unitary evolution.  

\emph{The model of a nuclear spin bath.---} Our discussion of information recovery is applicable to all systems that show quantum information scrambling \footnote{Supplementary Material provides an additional numerical proof of our predictions for the scrambling unitaries that were constructed as quantum circuits made of randomly generated two-qubit unitary gates.}. Thus, this effect should be possible to demonstrate with a processor that was used in the quantum supremacy test \cite{Qsupremacy}, as well as in simulations of the SYK model \cite{Kitaev2015,Sachdev1993-ka} driven according to the protocol in Fig.~1.  

As an example of that the Haar random unitary provides a correct description of the  behavior of OTOCs, we studied  numerically the nuclear spin-bath model. This model has been frequently used to describe interactions of a solid state qubit with nuclear spins \cite{nina2,bath-half1}. Its Hamiltonian is
\begin{equation}\label{eq:Hamiltonian}
    H = \sum_{i=1}^{N_s}\sum_{\alpha} J_i^{\alpha} S^\alpha s_i^\alpha, \ \alpha = x,y,z,
\end{equation}
where the couplings $J_i^{\alpha}$ are independent Gaussian distributed random numbers with zero mean and standard deviation $J$. Here, the central spin-$1/2$, $S$, interacts with the bath of $N_s$ spins-$1/2$'s, $s_i$. We simulated the Schr\"odinger equation with this Hamiltonian numerically to obtain the effect of the unitary evolution.

\begin{figure}[t!]
    \centering
    \includegraphics[width=0.45\textwidth]{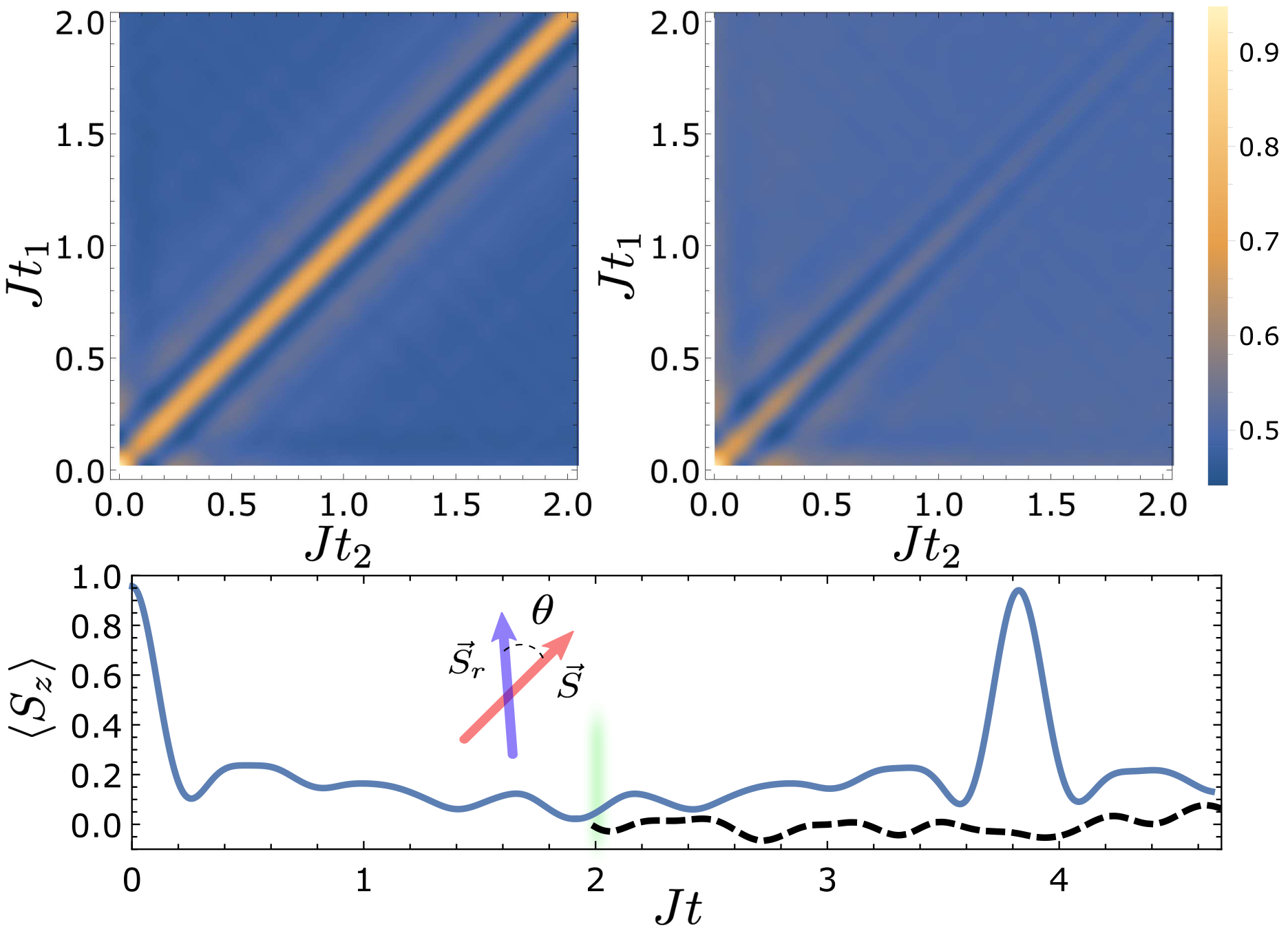}
    \caption{Top: Final measurement probabilities obtained from numerical simulations of the model (\ref{eq:Hamiltonian}) with quantum spins. Left and right panels correspond to the final probabilities deduced from the joint probability (\ref{eq:correlatorP}) and (\ref{eq:correlatorP2}), respectively. Bottom: Dynamics of the $z$ component of the  central spin vector  in model (\ref{eq:Hamiltonian}) with all classical spins. The green bar marks time $t_1$ with invasive measurement of the central spin. Black dashed curve and blue solid curve correspond to the cases with and without intermediate invasive measurement, respectively. 
    $N_s=10$ and $N_s=30$ for quantum and classical simulations, respectively.}
    \label{fig:density}%
\end{figure}

In our simulations, the central spin starts from the $|\uparrow\rangle$ state, i.e., $\sigma_i = \sigma_z$ in Eq.~(\ref{eq:finalP}). The bath spins are prepared in a maximally mixed state. This corresponds to a common situation of practically infinite temperature for nuclear spins.  Figure~\ref{fig:density}~(top~left) shows the obtained final Alice's probability of a nonzero measurement result for $\sigma_f=\sigma_z$ and a fixed randomly chosen $\sigma_r$.  It does show the echo effect along the $t_1=t_2$ line, in perfect quantitative agreement with Eq.~(\ref{eq:finalP}). No additional averaging over Bob's measurement axes and parameters $J_{i}^{\alpha}$  is needed for a quantitative agreement with Eq.~(\ref{eq:finalP}). We also verified numerically  for $N_s=10$ (not shown) that the same universal result is obtained for several other choices of the initial density matrix of the spin bath.

It is instructive to compare the echo in OTOC with another type of spin echo, which is induced by a similar protocol, in which the backward evolution $\hat{U}^\dag(t_2)$ is replaced by the forward one $\hat{U}(t_2)$. The joint probability is then described by the correlator~(\ref{eq:correlatorP}) with $t_2$ replaced by $-t_2$, i.e.,
\begin{equation}\label{eq:correlatorP2}
    \text{Prob}_2(\hat{P}_f,\hat{P}_r)=\langle \hat{P}_r(t_1)\hat{P}_f(t_1+t_2)\hat{P}_r(t_1)\hat{P}_i\rangle.
\end{equation}
This correlator can be measured in solid state systems by standard means, for example, it was studied experimentally in semiconductor quantum dots \cite{bechtold-prl} (see also Ref.~\cite{Liu-corr}). As we show in Fig.~\ref{fig:density}~(top right), the final probability (\ref{eq:correlatorP2})  also shows an echo effect near $t_2=t_1$. However, this echo originates from a finite scrambling rate and therefore decays at large times, in sharp contrast to the echo in the OTOC. 

{\it Quantum vs classical scrambling.---}  Our observation can be used to additionally validate quantum supremacy tests such as in Ref.~\cite{Qsupremacy}.  
There is only a finite depth for precise simulations of classical chaotic motion with classical digital computers because of the exponential increase of round-off errors \cite{chaos-truncation}. Therefore, a classical analog computer with nonlinear interactions between its components can evolve reversibly to a state that cannot be predicted with classical digital computers. Hence, to validate quantum supremacy, an additional test may be needed to prove that we deal with a quantum scrambled state at the end rather than with a state generated by classical chaos \cite{analog-supr}. 

In the classical case, a small change of the state vector at the end of the forward protocol would be quickly magnified during the backward time evolution. Thus, the state at the end of our control protocol would be strongly different from the initial one. 

\begin{figure}[t!]
   \centering
    \includegraphics[width=0.45\textwidth]{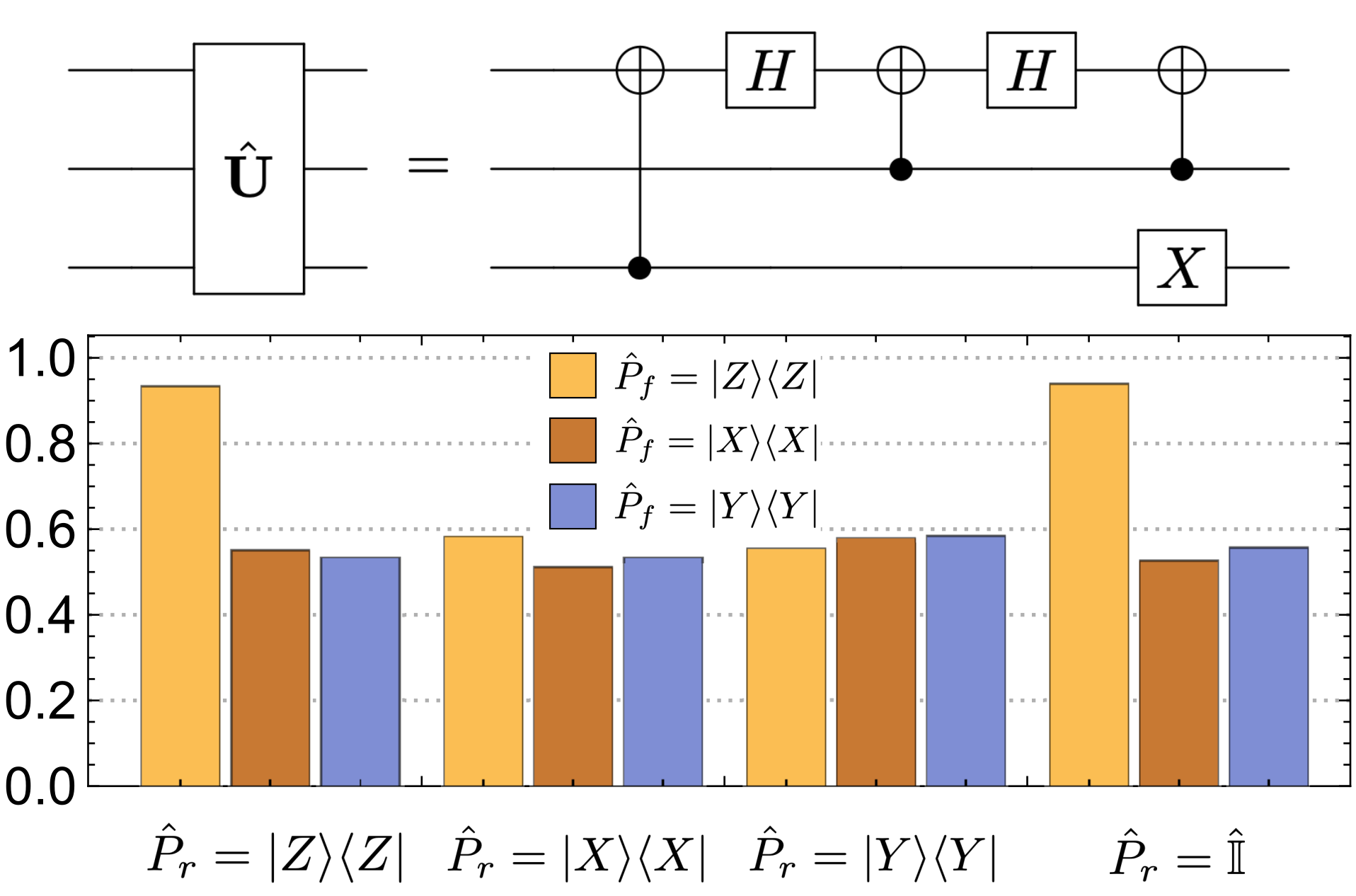}
    \caption{Top: The realization of the unitary (\ref{eq:unitary}). Bottom: Statistics of the measurement probabilities obtained with the IBM-Q processor. Colors label orthogonal final measurements of $\hat{P}_f$. The three groups of the chart correspond to orthogonal intermediate measurements $\hat{P}_r$. The basis is $|Z\rangle=|0\rangle$, $|X\rangle=(|0\rangle+|1\rangle)/\sqrt{2}$, and $|Y\rangle=(|0\rangle+i|1\rangle)/\sqrt{2}$.
    The final measurement probabilities, after averaging over intermediate outcomes, read as $0.5426$, $0.5522$, and $0.7529$, for $\hat{P}_f = |X\rangle\langle X|$, $|Y\rangle\langle Y|$, and $|Z\rangle\langle Z|$, respectively. }
    \label{fig:chart}
\end{figure}

To illustrate this, we simulated the evolution of interacting classical spins with the same Hamiltonian (\ref{eq:Hamiltonian}) subject to the classical Landau-Lifshitz equations \cite{Sinitsyn2012-fv}. The state of a classical spin is specified by a three dimensional unit vector. Our initial state for the central spin is a unit vector pointing along the $z$ axis. Bath spins start from random directions. Instead of quantum projective measurements, we assumed that classical measurements were invasive. Namely, a measurement resets the classical spin to be directed either along or opposite to the measurement axis with probabilities $\cos^2{(\theta/2)}$ or $1-\cos^2{(\theta/2)}$, respectively, where $\theta \in [0,\pi]$ is the angle between the central spin vector and the direction of the measurement axis. The time-reversed dynamics was induced by changing the sign of all spin coupling constants.

Figure~\ref{fig:density}~(bottom) shows the evolution of the $z$ component of the central classical spin during the  time of the protocol. As expected, an intermediate invasive measurement of only the central spin has lead then to a state with unrecoverable initial information, in sharp contrast to the quantum case.

\vspace{5pt}
\emph{IBM-Q experiment.---}
To verify that our predictions are robust against weak natural decoherence that is always present in modern quantum computers, we performed an experiment with the IBM-Q  five-qubit processor. The main programmed system consisted of one central and two bath qubits. We also added an ancillary qubit to effectively realize  measurement by entanglement of this and the central qubit. 

The quantum circuit is the same as in Fig.~\ref{fig:device}. Here, the three qubit unitary is given by
\begin{equation}\label{eq:unitary}
\begin{aligned}
        \hat{\bf{U}}=\mathbb{I}\otimes |01\rangle&\langle 00| + \sigma_x \otimes |00\rangle\langle 01| \\
        &-i\sigma_y\otimes |11\rangle\langle 10|-\sigma_z\otimes |10\rangle\langle 11|,
\end{aligned}
\end{equation}
with the corresponding quantum circuit realization shown in Fig.~\ref{fig:chart} (top) \footnote{This unitary was used before to demonstrate the no-hiding theorem on the quantum computers \cite{Kalra2019-jy}.}. It maps any initial state $|i\rangle$ of the target qubit to the maximally mixed state, provided that the input states of the bath qubits are $|+\rangle=(|0\rangle+|1\rangle)/\sqrt{2}$. 
This unitary is not a typical Haar random unitary, since its scrambling effect depends on the bath's initial state. Consequently, the four point correlator at $t_2=t_1$ in Eq.~(\ref{eq:jointP}), i.e., $\langle \hat{U} \hat{\sigma}_r \hat{U}^\dag \hat{\sigma}_f \hat{U}\hat{\sigma}_r\hat{U}^\dag\hat{\sigma}_i \rangle$, is not zero, even after averaging $\hat{\sigma}_r$ over random Pauli matrices (corresponding to random intermediate projective measurements). Instead, it makes the above four point correlator vanish when $\hat{\sigma}_r$ is averaged over the Pauli group. Thus, if we include the identity operator as part of the intermediate measurement, the rest of the protocol and the final result will not change.

The central qubit starts at $|0\rangle$. We perform the final measurements along three orthogonal directions, corresponding to $|0\rangle$, $(|0\rangle+|1\rangle)/\sqrt{2}$, and $(|0\rangle+i|1\rangle)/\sqrt{2}$. The theoretical probabilities for these final measurements are $0.75$, $0.5$, and $0.5$, respectively. Figure~\ref{fig:chart} shows the statistics of the final measurements, obtained with the IBM $5$-qubit processor, from which we inferred   the initial state  $0.992|0\rangle-(0.082-0.101i)|1\rangle$. This corresponds to $0.983$ fidelity, which means that natural decoherence was not detrimental.

The effect of damaged information recovery should be accessible for verification in more complex systems using capabilities in the generation of random unitary evolution \cite{Qsupremacy,Li2017-il,Samal2011-sm}. In addition, we showed that this effect is described by the long-time saturation values of OTOCs. This observation reveals a novel domain of applications for such unusual correlators.

This effect becomes very counterintuitive if we  interpret unitaries $U$ and $U^{\dagger}$ in Fig.~1 as, respectively, backward and forward time travel operators.  Then, the intermediate Bob's measurement is expected to lead to the same butterfly effect as the one in the famous  Ray Bradbury's story ``A Sound of Thunder'' \cite{bradbury}.  In that story, a character used a time machine to travel to the deep past, stepped on an insect there, and  after returning to the present time found a totally different world.
In contrast, our result shows that by the end of a similar protocol the local information is essentially restored.

\begin{acknowledgements}
We would like to thank Avadh Saxena and Wojciech Zurek for comments and suggestions. This work was supported by the U.S. Department of Energy, Office of Science, Basic Energy
Sciences, Materials Sciences and Engineering Division, Condensed Matter Theory Program.
\end{acknowledgements}

\bibliography{reference}

\clearpage
\appendix

\setcounter{page}{1}
\renewcommand\thefigure{\thesection\arabic{figure}}
\setcounter{figure}{0} 

\onecolumngrid

\begin{center}
\large{ Supplementary Material for \\ ``Recovery of Damaged Information and the Out-of-Time Order Correlators''
}
\end{center}

\section{Random Circuit Model}

Here, we show the results of our numerical study of information scrambling in a quantum circuit model composed of random two-qubit gates. This provides a different, from the nuclear spin model, illustration of the information recovery effect, at conditions that mimic computations by the Google's processor for testing quantum supremacy.
We also explore here the deviations from our theoretical predictions due to the finite system size effects.

The probability for the final measurement $\hat{P}_f$ (at $t_2=t_1$), after averaging over random scrambling unitaries, was shown (Eqs.~(10) and (11) in the main text) to be
\begin{equation}\label{eq:supp_final}
    \text{Prob}(\hat{P}_f)=\frac{1}{2}+\frac{1}{8}\langle \hat{\sigma}_f\hat{\sigma}_i \rangle + \bar{F},
\end{equation}
where the term $\bar{F}$ is the Haar average of an OTOC, i.e.,
\begin{equation}
    \bar{F} = \int_{Haar} dU\ \tr \left[ U^\dag\sigma_rU \sigma_i U^\dag\sigma_rU \sigma_f \rho_B \right] =\tr(\sigma_i\sigma_f\otimes\rho_B)/(N^2-1),
\end{equation}
where $N$ is the dimension of the Hilbert space. Hence, $\bar{F}$ vanishes rapidly with increasing system size.

We claim that for relatively large system size a single typical scrambling unitary is representative, i.e., it gives a result that agrees with $\bar{F}$ up to negligible finite size fluctuations.
Generally, it is sufficient to show for a  large number of qubits  that the deviation between the typical OTOC value and $\bar{F}$ is vanishing.  This deviation is characterized by 
\begin{equation}\label{eq:variance}
 C = \int_{Haar} dU\ \left( [\text{Prob}_U(\hat{P}_f)]^2 - [\text{Prob}(\hat{P}_f)]^2 \right).
\end{equation}
Here $\text{Prob}_U(\hat{P}_f)$ is the measurement probability for a single scrambling unitary $U$. 
We expect that fluctuations  decrease with the system size as,
\begin{equation}
    C \sim N^{-1},
\end{equation}
where $N$ is the size of the phase space $N=2^{n_q}$, and where $n_q$ is the number of qubits.

\begin{figure}[h!]
    \centering
    \includegraphics[width=0.4\textwidth]{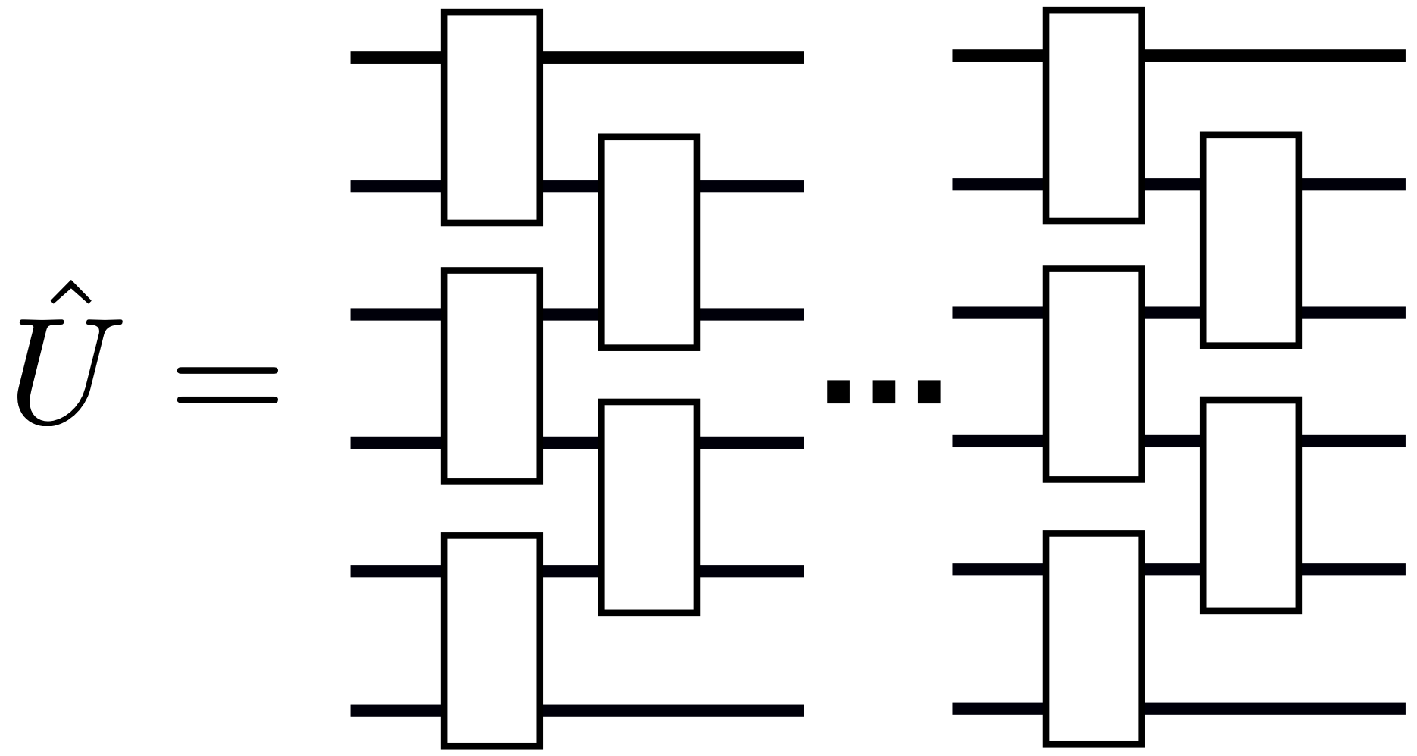}
    \caption{The global scrambling unitary is composed of multiple layers of random two-qubit gates. Each two-qubit gate is drawn from the unitary group with respect to Haar measure.}
    \label{fig:twounitary}
\end{figure}

To test this prediction, 
we  simulated numerically the dynamics induced by a quantum circuit composed of random two-qubit unitaries. Namely, the global random scrambling unitary in the protocol is composed of random two-qubit unitaries, as illustrated in Fig.~\ref{fig:twounitary} of this supplementary. In our simulations, each scrambling unitary was generated with $1000$ layers of such two qubit unitaries. All input states of the qubits were prepared in computational $|0\rangle$.

\begin{figure}[h!]
    \centering
    \includegraphics[width=0.55\textwidth]{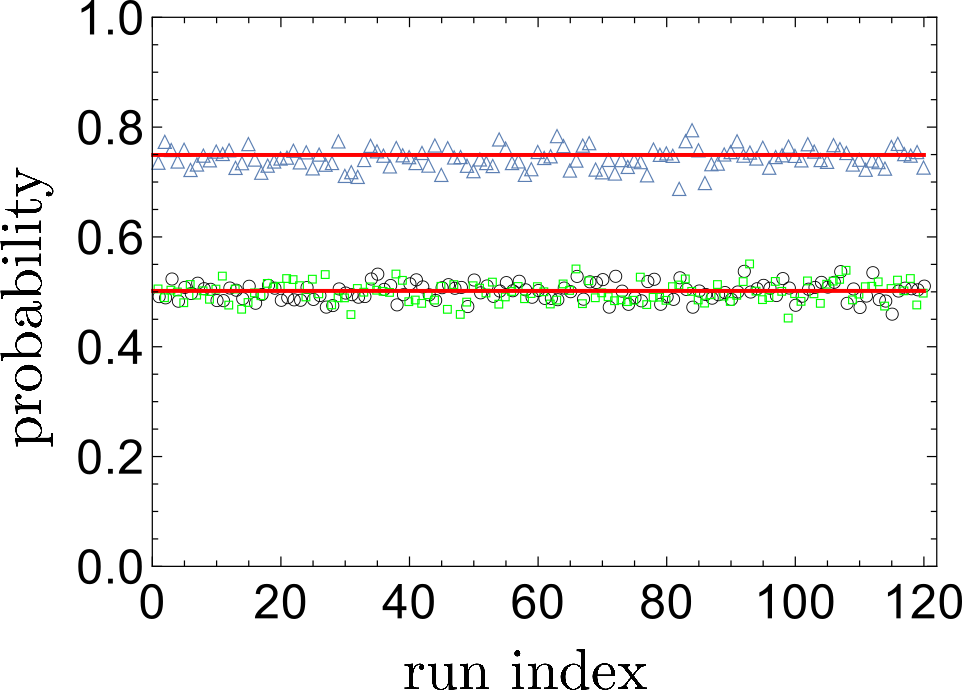}
    \caption{ Final Alice's measurement probabilities for many individual runs, with different random unitary evolution.  Blue triangles, green squares, and black circles correspond to final projective measurements $|Z\rangle\langle Z|$, $|X\rangle\langle X|$, and $|Y\rangle\langle Y|$, respectively, where $|X\rangle = |0\rangle$, $|Y\rangle = (|0\rangle+|1\rangle)/\sqrt{2}$, and $(|0\rangle+i|1\rangle)/\sqrt{2}$.  Red and orange lines are the theoretically predicted values.}
    \label{fig:runs}
\end{figure}

Figure~\ref{fig:runs} shows the final numerically obtained measurement probabilities, $\text{Prob}(\hat{P}_f)$ for different single runs of the protocol in Fig.~1 of the main text for eight qubits. The results show very small fluctuations, for each randomly chosen unitary, around the theoretically predicted mean values. 

\begin{figure}[h!]
    \centering
    \includegraphics[width=0.6\textwidth]{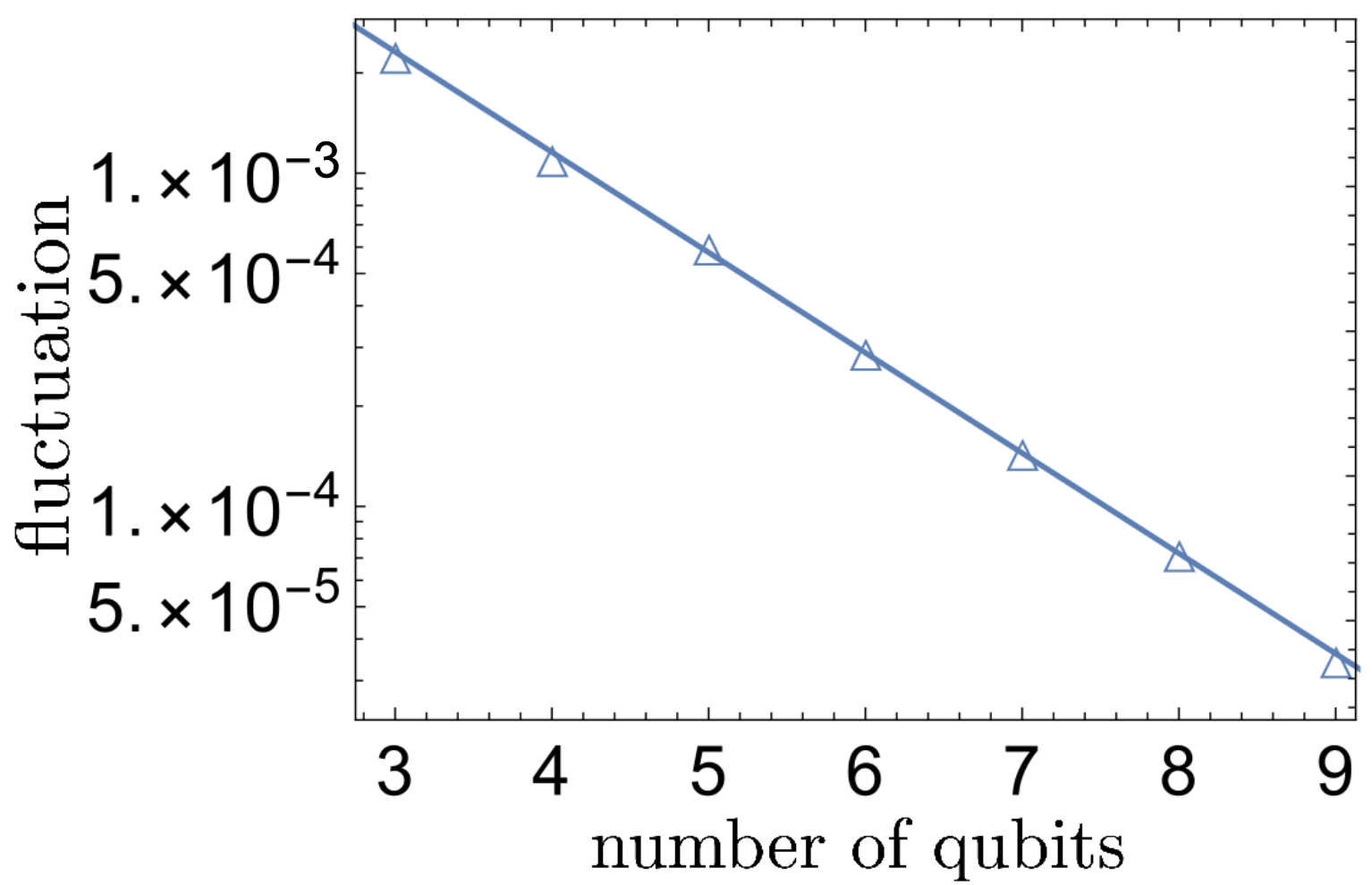}
    \caption{Fluctuation variance (\ref{eq:variance}) for different numbers of qubits. Triangles correspond to numerically obtained values for the variance. The solid curve is the fit by the $\sim e^{-n_q}$ curve, where $n_q$ is the number of qubits. This proves that deviations from our theoretical predictions are suppressed with the number of qubits exponentially quickly.}
    \label{fig:fluct}
\end{figure}

By measuring such deviations for different numbers of qubits, we studied the fluctuation scaling with the increasing system size. Fig.~\ref{fig:fluct} shows the numerically extracted fluctuations for different numbers of qubits in the system. The scaling is well described by $\sim N^{-1}$. 

\end{document}